\documentclass[a4paper,fleqn,usenatbib]{mnras}

\usepackage[T1]{fontenc}
\usepackage{ae,aecompl}

\usepackage{graphicx}	
\usepackage{amsmath}	
\usepackage{amssymb}	
\usepackage{wasysym}
\usepackage{stmaryrd}

\title{On extreme transient events from rotating black holes and their gravitational wave emission}

\author[van Putten and Della Valle]{Maurice H.P.M. van Putten$^{1,2}$ and Massimo Della Valle$^{3,4}$\thanks{E-mail: mvp@sejong.ac.kr} \\
$^{1}$Room 614, Sejong University, 98 Gunja-Dong Gwangin-gu, Seoul 143-747, Korea \\
$^{2}$Kavli Institute for Theoretical Physics, University of California, Santa Barbara, CA 93106-4030 USA \\
$^{3}$ Instituto Nazionale di AstrofisicaOsservatorio Astronomico di Capodimonte, Salita Moiariello 16, 80131 Napoli \\
$^{4}$ International Center for Relativistic Astrophysics, Piazzale della Repubblica 2, 65122 Pescara, Italy}

\date{Accepted XXX. Received YYY; in original form ZZZ}

\pubyear{2016}

\begin{document}
\label{firstpage}
\pagerange{\pageref{firstpage}--\pageref{lastpage}}
\maketitle

\begin{abstract}
	The super-luminous object ASASSN-15lh (SN2015L) is an extreme event with a total energy $E_{rad}\simeq 1.1\times 10^{52}$ erg in black body radiation on par with its kinetic energy $E_k$ in ejecta and a late time plateau in the UV, that defies a nuclear origin. It likely presents a new explosion mechanism for hydrogen-deprived supernovae. 
	{With no radio emission and no H-rich environment 
		we propose to identify $E_{rad}$ with dissipation of a baryon-poor outflow in the optically thick remnant stellar envelope produced by a central engine.} By negligible time scales of light crossing and radiative cooling of the envelope, SN2015L's light curve closely tracks the evolution of this engine. We here model its light curve by the evolution of black hole spin, during angular momentum loss in Alv\'en waves to matter at the Inner Most Stable Circular Orbit (ISCO). 
	{The duration is determined by $\sigma=M_T/M$ of the torus mass $M_T$ around the black hole of mass $M$: $\sigma\sim 10^{-7}$ and $\sigma\sim 10^{-2}$ for SN2015L and, respectively, a long GRB. The observed electromagnetic radiation herein represents a minor output of the rotational energy $E_{rot}$ of the black hole, while most is radiated unseen in gravitational radiation. {This model explains the high-mass slow-spin binary progenitor of GWB150914, as the remnant of two CC-SNe in an intra-day binary of two massive stars.} This model rigorously predicts a change in magnitude $\Delta m\simeq 1.15$ in the light curve post-peak, in agreement with the light curve of SN2015L with no fine-tuning.}
\end{abstract}


\section{Introduction}

By a total energy $E_{rad}\simeq 1.1\times 10^{52}$ erg in black body (BB) radiation, ASASSN-15lh \citep[SN2015L][]{don16} is the brightest object so far among hydrogen-poor super-luminous supernovae (SLSNe-I) \citep{pas07,qui11,gal12}. The origin of its luminosity is an open question, because  the light curves of these SNe do not seem  powered by nuclear decay of $^{56}$Ni \citep{koz16}, common to conventional core-collapse supernovae \citep{pas10,wan15}. 
{Importantly, interactions with a circumstellar environment appear wanting, based on radio \citep{koo15} and optical spectroscopy \citep{mil15}.}

SN2015L may represent a radically new class of SLSNe by turning convention on its head with 
\begin{eqnarray}
E_{rad} \simeq E_k,
\label{EQN_rk}
\end{eqnarray}
where $E_k\sim 10^{52}$ erg denotes the kinetic energy in the envelope expanding at a few percent of the velocity of light, that derived from energy dissipation in an expanding envelope powered by a long-lived inner engine alone. Furthermore, its light curve features a late time plateau about two months post-peak with a change in magnitude
\begin{eqnarray}
\Delta m \equiv m_{pl} - m_p \simeq 1.2.
\label{EQN_Dpl}
\end{eqnarray}

To explain SN2015L, natural candidates for a long-lived inner engine are an angular momentum-rich magnetar \citep[e.g.][]{ins13,maz14} or black hole-disk system driving the explosion by baryon poor outflows \citep[e.g.][]{bis70}. Since the light crossing time scale of the expanding envelope of radius $\sim 5\times 10^{15}$ cm and its radiation cooling time are merely one day, SN2015L offers an interesting prospect for essentially real-time tracking of an evolving central engine, heating the remnant stellar envelope by a baryon-poor jet (BPJ).

Magnetars and rotating black holes have been widely considered as candidate inner engines of long gamma-ray bursts (LGRBs). While there durations $T_{90}$ of tens of seconds associated with relativistic core-collapse supernovae \citep[e.g.][]{woo06,fry15} are shorter than the light curve of SN2015L by a factor of about $10^5$, their true output in gamma-rays $E_\gamma\simeq 9\times 10^{51}$ erg \citep{fra01,ghi06,ghi13} is essentially equal to $E_{rad}$. This points to a possibly common energy reservoir, operating across a broad range of time scales with different modes of dissipation. While $E_\gamma$ is largely non-thermal and probably derives from internal shocks in ultra-relativistic baryon poor jets (BPJ) \citep{ree92,ree94,eic93,eic11}, 
SN2015L features essentially thermal electromagnetic radiation satisfying
\begin{eqnarray} 
E_{rad}\simeq E_{\gamma}.
\label{EQN_rg}
\end{eqnarray}
Attributing $E_{rad}$ to dissipation of a BPJ in the optically thick remnant stellar envelope, SN2015L represents a ``failed GRB" scaled down in luminosity and inverse duration. 

A one-parameter scaling across a broad range of luminosites and time scales exists in shedding Alv\'en waves in relativistic outfows from a common energy reservoir in angular momentum, operating at like efficiencies in conversion to electromagnetic radiation. For SN2015L, $L_{rad}=E_{rad}/T\simeq 2\times 10^{45}$ erg s$^{-1}$ and the duration $T$ of about two months post-peak to aforementioned plateau satisfy 
\begin{eqnarray}
\alpha = \frac{L_{rad}}{L_\gamma}\simeq \frac{T_{90}}{T} \simeq 10^{-5},
\label{EQN_alpha}
\end{eqnarray} 
where $L_\gamma\simeq E_\gamma/T_{90} \simeq 3 \times 10^{50}$ erg s$^{-1}$ for LGRBs of durations $T_{90}$ of tens of seconds. Even though SN2015L and LGRBs have principally different spectra, they may be operating at radiation efficiencies that are not too different.

Various authors attribute SN2015L to magnetars \citep{ber16,don16,dai16,suk16,cha16} as a variant to magnetar powered luminous type Ic supernovae \citep{wan15}, whose energy budget approaches the extremal value $E_c\simeq 7.5\times 10^{52}$ \citep{hae09} (but see \cite{met15}) at high efficiency \citep[e.g.][]{don16}. This scenario is challenged by some properties exhibited by the SN2015L event. Firstly, the similarity between the energy budget ($E_{rad}$+$E_k$) of SN 2015L and the extremal energy value provided by a rotating neutron star implies values for the energy conversion factors, close to 100\%, which appears unlikely. Secondly, SN2015L shows a late time plateau, not seen in conventional core-collapse supernovae \citep[e.g.][]{tur90} nor in broad lined SNe Ic \citep{buf12}. The algebraic time decay $\propto (1 + t/t_s)^{-2}$ of magnetar light curves with characteristic spindown time $t_s$ is woefully at odds with aforementioned unconventional temporal behavior. 
{Thirdly, versions of this model that address the UV plateau invoke interaction of the envelope with the circumstellar environment, which is at odds with above cited radio non-detection and optical non-detection of H$\alpha$ emission. Being hydrogen deficient deprives the expanding envelope in SN2015L to interact energetically with a dense pre-supernova wind (cf. super bright SNe hydrogen rich explosions).} {\em SN2015L hereby points to a long-lived central engine with inherent decay to a late-time plateau.}

{Moreover, amongst the SLSNe-I class, SN2015L is the brightest observed thus far.}	It is likely that even brighter events exist satisfying $E_{rad} >  E_c$. It is not without precedent that events discovered in our ``local" universe represent the faint tail of a broader class of astrophysical objects. For instance, SN1987A \citep{bur87}, SN1885A \citep{vau85} or GRB980425 \citep[but not the accompanying SN1998bw;][]{gal98} 
are certainly not the brightest members in their respective classes. 

{In light of the qualitative and quantitative considerations above, we here consider the possibility that SN2015L is powered by black hole-disk system, whose light curve tracks the evolution and state of accretion of the latter.} According to the Kerr metric \citep{ker63}, the rotational energy $E_{rot}$ can reach up to 29\% of its total mass, which exceeds that of neutron stars by well over an order of magnitude. A rotating black holes of mass $M$ hereby possess rotational energies up to
\begin{eqnarray}
E_{rot} \simeq 6 \times 10^{54}\,\left(\frac{M}{10 M_\odot}\right)
\,\mbox{erg}.
\label{EQN_rH}
\end{eqnarray}
in the limit as the dimensionless Kerr parameter $a/M$, defined by the ratio of specific angular momentum $a$ to black hole mass, approaches unity. Near-extremal black holes with $a/M=0.9$ hereby have about one-half the maxima rotational energy of a black hole with the same mass energy-at-infinity $M$. Rapidly rotating stellar mass black holes hereby provide ample energies for SN2015L and even brighter events.

$E_{rad}$ in SN2015L and $E_\gamma$ in LGRBs represent about 1\% of $E_{rot}$ - a {\em very minor fraction} of the total energy reservoir of a Kerr black hole. Attributing it to a fraction of horizon magnetic flux, $E_{rad}$ derives from the total energy $E_{BPJ}$ along an open magnetic flux tube subtended by a horizon half-opening angle $\theta_H$ satisfying
\begin{eqnarray}
\frac{E_{BPJ}}{E_{rot}} = \frac{1}{4}\theta_H^4<<1,
\label{EQN_1p}
\end{eqnarray} 
supported by an equilibrium magnetic moment of the black hole (\S4, below). Should a major fraction of (\ref{EQN_rH}) be released, it must be largely so in channels unseen, i.e., gravitational radiation and MeV neutrinos. These fractions are determined by the partition of Alfv\'en waves around the black hole \citep{van08b,van15b}, producing a BPJ along the black hole spin axis and interactions with nearby matter at thet the Inner Most Stable Circular Orbit (ISCO) \citep{van99,van03}. These Alfv\'en waves are induced by relativistic frame dragging. The existence of frame dragging is not in doubt: recent measurements of non-relativistic frame dragging around the Earth are in excellent agreement with general relativity \citep{ciu04,ciu07,ciu09,eve11}. 

This partition (\ref{EQN_1p}) applies to a state of suspended accretion. Earlier models envision outflows from accreting black holes \citep{ruf75,bla77}. It has been suggested that the BPJ represents the major energetic output from the black hole in the form of a Poynting flux \citep{bla77}. Below a critical accretion rate \citep{glo14}, however, the black hole may act back onto matter at the ISCO and sufficiently so to suspend accretion \citep{van99,van03}. Thus, different evolutions of the black hole ensue depending on the state of accretion: {\em spin up}, described by modified Bardeen accretion \citep{van15b}, and {\em spin down}, by equations of suspended accretion \citep{van03}. The latter introduces a radically new secular time scale, defined by the lifetime of rapid spin of the black hole \citep{van03}
\begin{eqnarray}
T_{s}\simeq 100 \left(\frac{\sigma}{10^{-7}}\right)^{-1}\,\mbox{day},
\label{EQN_Ts}
\end{eqnarray}
defined by the ratio $\sigma=M_T/M$ of the mass $M_T$ of a torus at the ISCO to $M$. This process features a {\em minor} output (\ref{EQN_1p}) in a BPJ accompanied by a {\em major} output in gravitational radiation from matter at the ISCO. The latter features a distinctive descending chirp \citep{van08a}. 

By (\ref{EQN_1p}-\ref{EQN_Ts}), our model is different from others aforementioned in the partition of energy output - mostly into surrounding matter accompanied by minor emisison in open outflows - and a new secular timescale in the lifetime of black hole spin, that may extend well beyond canonical time scales of accretion.

Different modes of accretion naturally appear in black hole evolution, marked by distinct total energies and light curves for frame-dragging induced BPJs. They same applies to any accompanying emissions unseen in gravitational waves from matter at the ISCO, that may be probed by LIGO-Virgo and KAGRA \citep{abr92,ace06,ace07,som12,lig16}.	
For LGRBs, a model light curve for BPJ during the final phase of spin down \citep{van03} is supported by spectral-energy and temporal properties on long and short time scales \citep[summarized in][]{van16} in GRBs  detected by HETE-II and {\em Swift} \citep{swi04,swi14}, {\em Burst and Transient Source Experiment} (BATSE) \citep{kou93} and {\em BeppoSAX} \citep{fro09}, whose durations of tens of seconds has been association with high density accretion flows \citep{woo93,mac99,woo06,woo10}. Conceivably, this light curve also has relevance to {failed GRBs}, i.e., when the BPJ fails to break out of the progenitor stellar envelope. 

{By the above, we associate SN2015L and long GRBs to scaling (\ref{EQN_alpha}) with the torus to black hole mass ratio $\sigma$ according to $\sigma\simeq 10^{-7}$ and, respectively, $\sigma\simeq 10^{-2}$ in (\ref{EQN_Ts}). {Our $\sigma$ in (\ref{EQN_Ts}) is intermediate in being a geometric of $\sigma\simeq10^{-2}$, which is typical for GRBs, and $\sigma\simeq 10^{-12}$ in the micro-quasar GRS1915+105 \citep{mir94,mil16} and long GRBs.}
	
	To model the associated transient emission, we first consider the phases of black hole evolution by fallback matter from a rotating progenitor following birth in core-collapse (\S2) and their association with supernovae (\S3). We then revisit model light curve for transient emission during accretion and and spin down (\S4). In \S5-6 we give a detailed confrontation of our model light curves with LGRBs and SN2015L, the former against data from BATSE, {\em BeppoSAX}, {\em Swift} and HETE-II. A summary and an outlook on accompanying gravitational wave emission is given in \S7.}

\section{Black hole evolution in core-collapse}

By core-collapse, massive stars are believed to be factories of neutrons stars and
black holes. In the absence of any direct observation by gravitational radiation, the details of black hole birth remain uncertain. It may form by prompt collapse. For instance, at the onset of core-collapse, fall back matter can form highly asymmetric distributions at centrifugal hang-up, producing a brief flash of gravitational waves \citep{due04}. By gravitational radiation and diluting angular momentum during continuing fallback, this state should be short-lived. If not, rather exotic objects might form \citep{lip83,lip09}. Alternatively, it forms through core-collapse of a neutron star, and similar flash in gravitational waves might be produced \citep{ree74}. 

Regardless of details at birth, the newly formed black hole must satisfy the Kerr constraint 
\begin{eqnarray}
\frac{a}{M} \le 1,
\label{EQN_KC}
\end{eqnarray}
where $a=J/M$ denotes the specific angular momentum of a black hole of mass $M$. Here, we use geometric units in which Newton's constant $G$ and the velocity of light $c$ are set equal to 1. In a uniformly rotating core, a newly formed black hole form near-extremal and of small mass \citep[e.g.][]{van04}. It then enters the following three phases in evolution.
\begin{enumerate}
	\item Phase I. {\em Surge to a non-extremal black hole.} The black hole increases its mass in direct accretion of a fraction of the inner region of the core, whose specific angular momentum is insufficient to stall at the ISCO. The dimensionless Kerr parameter $a/M$ hereby decreases on the free fall time scale of tens of seconds of the progenitor;
	\item Phase II. {\em Growth to a near-extremal black hole.} Accretion continues from the ISCO on the viscous time time scale of a newly formed accretion disk, increasing $M$ {\em and} $a/M$ \citep{bar70}, also in the presence of outflows at typical efficiencies \citep{van15b}. The black hole becomes near-extremal ($a/M\simeq 1)$ provided there is sufficient fall back matter and/or the
	progenitor rotates sufficiently rapidly;
	\item Phase III. {\em Spin down to slow spin.} The black hole looses decreases in $a/M$ by angular momentum loss against matter at the ISCO mediated by an inner torus magnetosphere supported by a torus about the ISCO, provided that (a) there is continuing fall back matter (a near-extremal black hole formed in Phase II) and (b) the accretion rate is subcritical \citep{glo14,van15b}. The duration of this suspended accretion state depends on the energy in the poloidal magnetic field, bounded by of a torus at the ISCO \citep{van03}.
\end{enumerate}

\begin{figure}
	\centerline{\includegraphics[scale=.5]{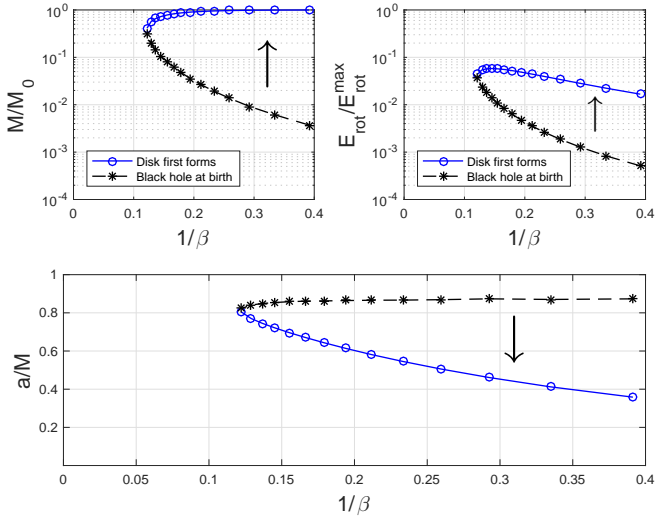}}
	\caption{Formation of initially extremal black hole in prompt core-collapse from a rotating progenitor, satisfying the Kerr limit $a/M=1$ and its subsequent surge by direct accretion of fall back matter to a non-extremal black hole, until a disk first forms. The results depend strongly on the angular velocity of the progenitor, here parameterized by the dimensionless period $\beta$ in (\ref{EQN_bet}). (Adapted from \citep{van04}).}
	\label{FIG1a}
\end{figure}

\begin{figure}
	\centerline{\includegraphics[scale=.5]{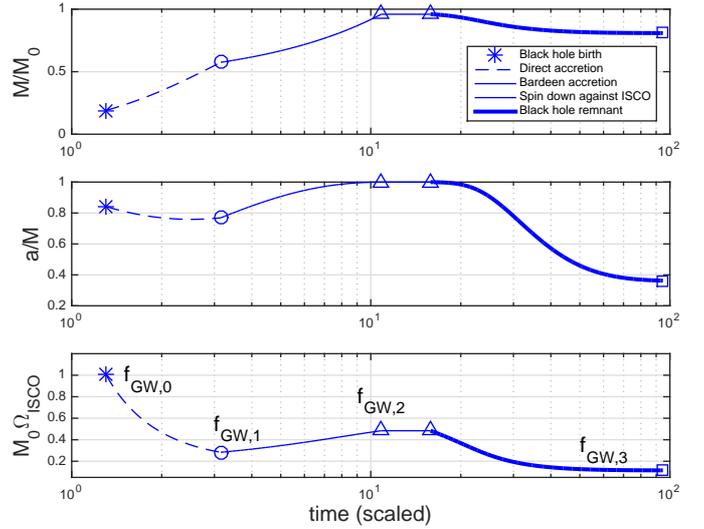}}
	\caption{Evolution of a rotating black hole following birth in a progenitor of mass $M_0$ in three phases of accretion: surge in direct accretion ({\em dashed}), growth by Bardeen accretion ({\em continuous}) followed by spin down against matter at the ISCO when accretion becomes subcritical (top and middle panels). Shown is further the associated evolution of any quadrupole gravitational wave signature from matter at the ISCO, marked by frequencies $f_{GW_0}$ at birth, $f_{GW1,}$ at the onset of Bardeen accretion, $f_{GW,2}$ at the onset of spin down and $f_{GW,3}$ at late times, when the black hole is slowly rotating in approximate corotation with matter at the ISCO (lower panel.)}
	\label{FIG1b}
\end{figure}

\begin{figure}
	\centerline{\includegraphics[scale=.5]{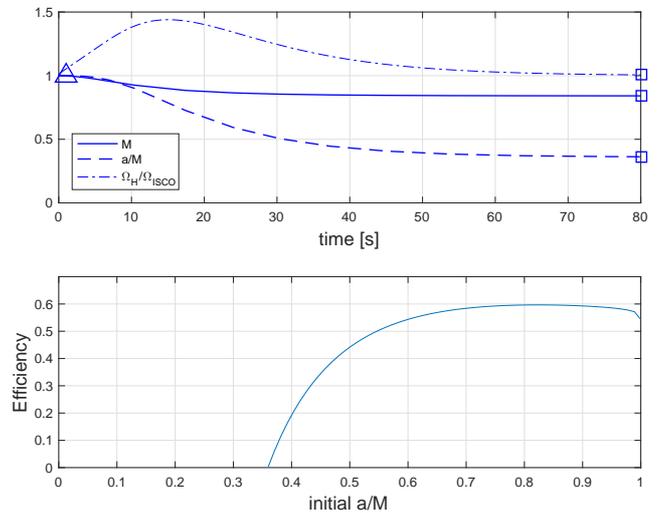}}
	\caption{In Phase III, spin down of a near-extremal black hole ({\em top}), formed in a prior phase of Bardeen accretion, represents catalytic conversion of spin energy into various emission channels by matter at the ISCO. {Shown are the evolution of total mass $M$ (solid curve), the dimensionless specific angular momentum $a/M$ (dashed curve) and ratio of angular velocities of the black hole event horizon to test particles at the ISCO (dot-dashed curve).} The efficiency ({\em bottom}) in energy extracted over initial rotational spin can reach 60\%.}
	\label{FIG1c}
\end{figure}

Phase I, considers direct accretion of low angular momentum from the inner
region of a progenitor rotating with period $P=P_d$ day. $P$ may be short, as when in cororation with a companion star in a compact stellar binary. We define the dimensionless reciprocal period $\beta$ \citep{van04}
\begin{eqnarray}
\beta = 4.22\,P_d^{-1} R_1^{-1} M_{He,10}^{-1/3},
\label{EQN_bet}
\end{eqnarray}
where $R=R_1 R_\odot$ ($R_\odot = 7\times 10^5$ km) and $M_{0}=M_{He,10}10M_\odot$ denotes the progenitor He mass.	Direct accretion of fallback matter falls occurs provided that its specific angular momentum is less than that at the ISCO, defined by the Kerr metric \citep{bar72,mac99,bet03,van04}. This condition is satisfied by a finite fraction of the core of the progenitor star, and defines an accretion time scale of at most the free fall time scale
\begin{eqnarray}
t_{ff} \simeq 30 \, \mbox{s} \, \left(\frac{M_{0}}{10M_\odot}\right)^{-\frac{1}{2}}  \left(\frac{r}{10^{10}\,\mbox{cm}}\right)^{\frac{3}{2}}.
\label{EQN_tff}
\end{eqnarray}
$M$ hereby surges rapidly to a fraction of order unity of the progenitor mass $M_0$, until an accretion disk first forms. At this moment, the black hole is non-extremal with a typically moderate Kerr parameter $a/M$ shown in Fig. \ref{FIG1a}.

In Phase II, the black hole evolves by matter plunging in gradually from the ISCO \cite{bar70,bar72,bet03,van04,kin06}. Matter is envisioned to migrates to the ISCO by large eddy turbulent viscosity \cite{sha73} in magneto-hydrodyamical stresses, such as may arise from a magneto-rotational instability (MRI \cite{bal91,haw91}). Accretion hereby advects magnetic flux onto the event horizon, that may lead to the formation of a BPJ \citep{bla77}. If so, the black hole evolves by modified Bardeen accretion \citep{van15b}, increasing $M$ and $a/M$ at canonical efficiencies. In hyper-accretion flows dominated by Reynolds stresses, the black hole satisfies the Bardeen integral \citep{bar70}
\begin{eqnarray}
zM^2=\mbox{const.}
\label{EQN_zM}
\end{eqnarray}
for the idealized limit of no outflows, shown in Fig. \ref{FIG1b}. The outcome is a high mass near-extremal Kerr black hole that may reach the Thorne limit \cite{tho74,sad11}. In attributing the accretion rate to an MRI induced viscosity, duration of phase II is expected to scale with the density of the accretion flow. 

In Phase III (Figs. \ref{FIG1b}-\ref{FIG1c}), the black hole evolves by an interaction with matter at the ISCO dominated by Maxwell stresses in Alfv\'en waves, defined by a system of two ordinary differential equations representing conservation of total energy and angular momentum. The system describes a black hole luminosity $L_H=-\dot{M}$ and torque $\tau = -\dot{J}$ on the inner face of matter at the ISCO \citep{van99,van16}, satisfying
\begin{eqnarray}
\dot{M}=\Omega_T\dot{J},~~\dot{J}=-\kappa e_k \left(\Omega_H - \Omega_T\right),
\label{EQN_CC}
\end{eqnarray}
where $\kappa$ is the net (variance) in (turbulent) poloidal magnetic field and kinetic energy $e_k$. Matter at the ISCO is hereby heated, driving non-axisymmetric instabilities giving rise to a dominant output in gravitational radiation \citep{van03}. The system (\ref{EQN_CC}) features two fixed points $\Omega_H=\Omega_T$ when the angular velocity of the black hole $\Omega_H$ equals that of matter at the ISCO: {\em unstable} at maximal spin ($\Omega_H=1/2M$) and {\em stable} at slow spin ($\Omega_H<<1/2M$). It points to a gradual evolution from fast to slow spin, i.e., a relaxation of the black hole spacetime to that of a slowly rotating black hole.  Given a stability bound on the total energy in the poloidal magnetic field, $\kappa$ proportional to the mass $M_T$ of a torus at the ISCO, we have \citep{van03,van15b}, more specifically than (\ref{EQN_Ts}), 
\begin{eqnarray}
T_s \simeq 30\,\mbox{s}\,\left(\frac{0.01}{\sigma}\right)\left(\frac{M}{7M_\odot}\right)\left(\frac{z}{6}\right)^4,
\label{EQN_Tspin}
\end{eqnarray}
where $z=r_{ISCO}/M$ denotes the radius of the ISCO relative to the gravitational radius $M$ of the black hole. $T_s\propto \sigma^{-1}$ shows that dilute mass concentrations at the ISCO result in lifetimes of the engine much longer than that of long GRBs, and hence
\begin{eqnarray}
\alpha \propto \sigma^{-1}.
\label{EQN_Ts2}
\end{eqnarray}
At constant poloidal magnetic field energy-to-kinetic energy in matter at the ISCO \citep{van03,bro06}, (\ref{EQN_Ts2}) defines a scaling of $L_H$ and $T_s$ conform (\ref{EQN_alpha}). Phase II and III hereby satisfy similar scalings of durations with density in accretion flows.

Illustrating the above, Fig. \ref{FIG1c} shows a black hole evolution to a high mass near-extremal state through the three phases of direct and Bardeen accretion followed by spin down, according to  (\ref{EQN_zM}-\ref{EQN_CC}). In this example, the angular velocity of the progenitor is relatively slow, giving rise to black hole mass close to but still less than the mass $M_0$ of the progenitor at the end of Phase II. Faster angular velocity give an earlier transition to Phase III of a lower mass $M$ of the black hole, yet still near-extremal in $a/M$.  

\section{Supernovae from rotating progenitors}

{Figs. 1-3 show the initial black hole mass and its subsequent evolution by direct accretion to strongly depend on rotation of the progenitor star.} In attributing an accompanying aspherical supernova explosion to winds or jets coming of the newly formed rotating inner engine \citep{bis70,mac99}, a successful explosion may ensue during the subsequent Bardeen phase and final spin down, but less likely so during the first phase of direct accretion. According to Figs. 1-2, the outcome of direct accretion critically depends on the dimensionless angular velicity $\beta$. Our treatment is hereby completely different black hole mass estimates in core-collapse of non-rotating isolated stars (our limit of $\beta=0$), see, e.g., \citep{heg03}. 

{However, growth to a rapidly spinning black hole by Bardeen accretion is rather insensitive to $\beta$, unless $\beta$ is so low such that an accretion disk never forms.}
Figs. 1-2 show the following two cases.

{\em Rapidly rotation progenitors} limit the initial mass $M$ of the black hole, so that all three phases of accretion can proceed. Rapid rotation, e.g., in intra-day stellar binaries, allows the formation of an extremal black hole $(a/M=1)$ at a mass $M$ that may be appreciably below the mass $M_0$ of the progenitor at the end of Bardeen accretion. This case leaves a finite amount of fallback matter to initiate black hole spin down, after accretion becomes sub-critical.

{\em Slowly rotating progenitors} \citep[e.g.][]{hir04,yoo15} imply long-duration direct accretion that may be followed by Bardeen accretion, leading to a high mass non-extremal black hole $(a/M<1$) with $M$ relatively close or equal to $M_0$. In this event, spindown in a suspended accretion phase is unlikely, as essentially all fallback has been exhausted.

{The case of rapid rotation is particularly likely to produce powerful winds and jets driving a supernova, more so than during direct and Bardeen accretion only for slowly rotating progenitors.
	SN2015L likely occured in a short period stellar binary, giving rise to high $\beta$ and a stripped hydrogen envelope (cf. \citep{pac98}). In this event, dependence on metallicity is expected to be minor, in contrast to the same by stellar winds from isolated stars \citep{heg03}.}

\section{Model light curves in Phase II--III}

The short time scale of direct accretion falls outside the scope of the months-long light curve of SN2015L. We therefore focus on Phase II and III, that, if present, can have long durations on secular time scales set by the density of accretion flow. {\em Their different modes introduce distinct magnetic field topologies and black hole evolution associated with spin up and, respectively, spin down, expressed in distinct light curves.}

In Phase II, the outflow tracks a black hole spinning up according to (\ref{EQN_zM}) using canonical scaling relations for BPJs based on total black hole horizon flux \citep{bla77,nat15a,nat15b}. At quipartition values of magnetic field and mass density, the latter scales with mass accretion rate $\dot{m}$. Model light curve of during Phase II hereby track $\dot{m}$, commonly considered as power laws in time, 
\begin{eqnarray}
L_{BPJ(t)}\propto t^{-n},
\label{EQN_LII}
\end{eqnarray}
e.g., at early high accretion rates $(n=1/2)$ or low accretion rates $(n\ge1)$ \citep{kum08a,kum08b,dex13}.

\begin{figure}
	\centerline{\includegraphics[scale=0.5]{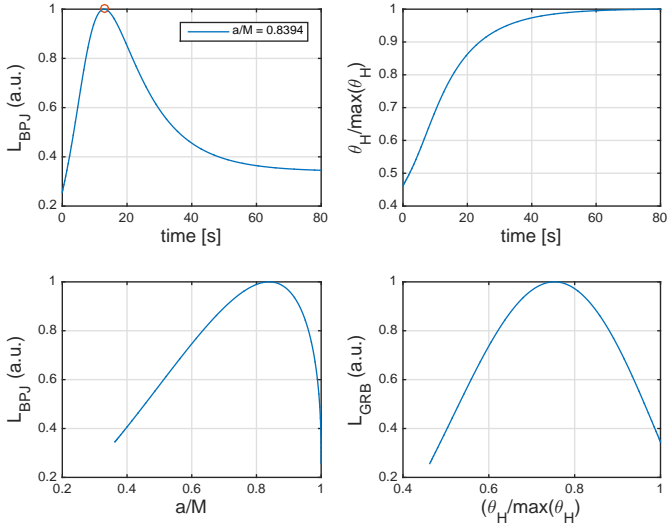}}
	\caption{Model light curve $L_{BPJ}$ of Phase III, produced along an open magnetic flux tube in a split field topology. It features a near-exponential decay post-peak at $a/M=0.8394$ to a plateau with change in magnitude $\Delta m= 1.154$. The horizon half-opening angle $\theta_H$ evolves to a constant.}
	\label{FIG2}
\end{figure}

In Phase III, the outflow tracks a black hole spinning down according to (\ref{EQN_CC}), wherein
$E_{BPJ}$ represent a {\em minor} fraction of $E_{rot}$ resulting in a FRED-like light curve \citep{van03}. A split magnetic field topology about the black hole in its lowest energy state \citep{van03,van15b} gives (\ref{EQN_1p}) by relativistic frame dragging along an open magnetic flux tube along the spin axis subtended by half-opening angle $\theta_H$ on the event horizon. 

To be specific, we consider \citep{van12} 
\begin{eqnarray}
L_{BPJ} \propto \hat{\theta}_{H}^4\Omega_H^2 z^n {\cal E}_k,
\end{eqnarray}
where $R_T=zM$ denotes the radius of the torus, ${\cal E}_k \propto (\Omega_TR_T)^2e(z)$ is the kinetic energy of the torus, and $e(z) = \sqrt{1-2/3/z}$ is the specific energy of matter in an orbit at angular velocity $\Omega_T$ at the ISCO \citep{bar72}. With $n=2$, $L_{BPJ}$ scales with the surface area within the torus and ${\cal E}_k$ is defined by a stability limit on poloidal magnetic field energy that the torus can support \citep{van03}. In a split magnetic flux topology, a maximal half-opening angle $\hat{\theta}_{H}\simeq 0.15$ rad accounts for a fraction of about 0.1\% of $L_H$ to be emitted along the spin axis of the black hole, the remaining 99.9\% being deposited into the torus for conversion into other radiation channels.

While at high spin $\theta_H$ appears to be correlated to $z$ \citep{van15b}, $\theta_H$ rises to a constant post-peak in the model light curve $L_{BPJ}(t)$. The outflow hereby effectively satisfies (\ref{EQN_alpha}). Post-peak, $L_{BPJ}(t)$ evolves by near-exponential decay in time to a plateau with a change in magnitude in $L_{BPJ}(t)$ satisfying
\begin{eqnarray}
\Delta m \equiv m_{pl} - m_p =1.154,
\label{EQN_Dm}
\end{eqnarray}
whose duration scales with $\sigma^{-1}$. This dimensionless number is a key model prediction,
that we shall compare with SN2015L (\S6 below).

\section{Observational tests on GRBs}

\begin{figure*}
	\centerline{\includegraphics[scale=0.47]{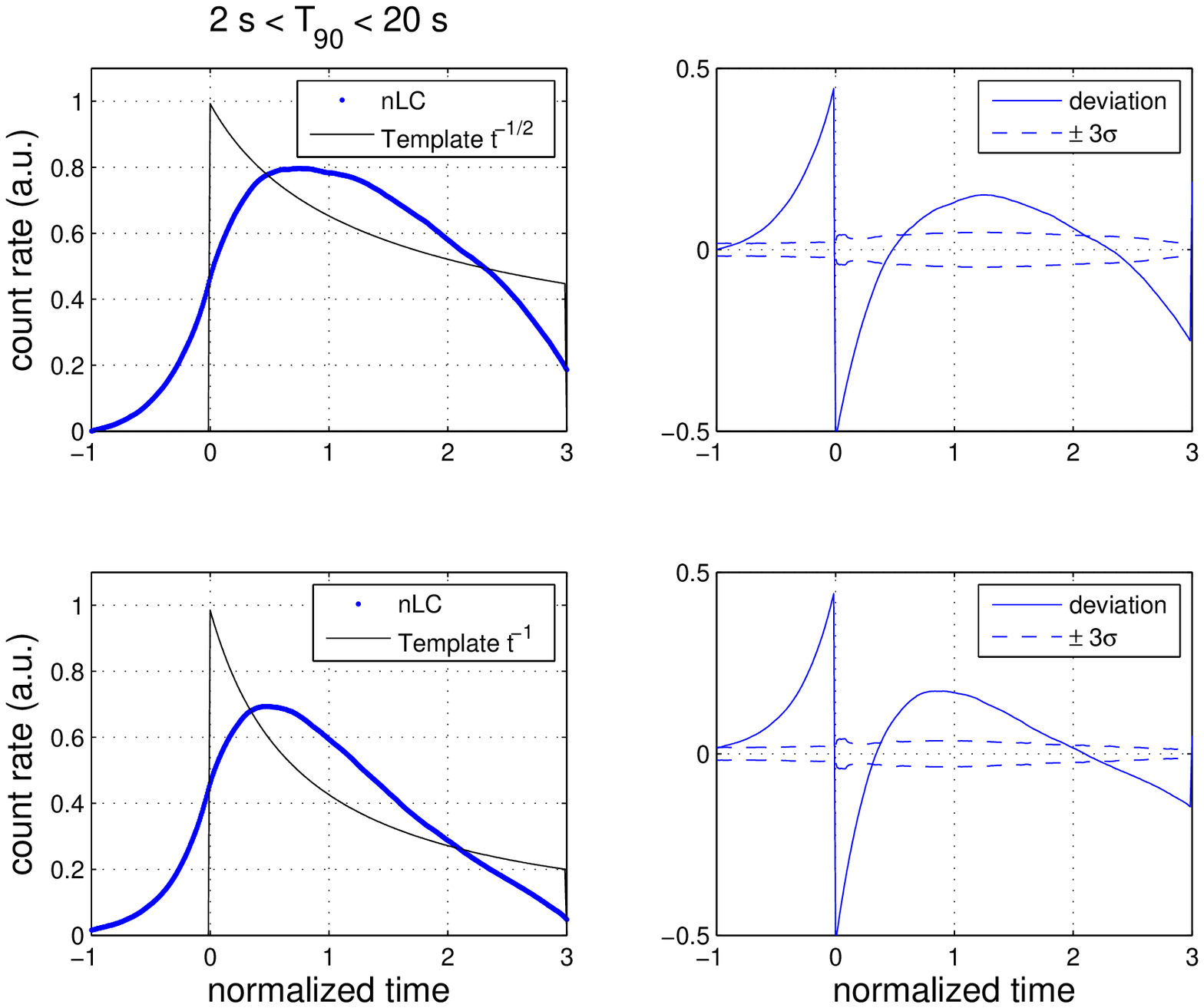}\includegraphics[scale=0.47]{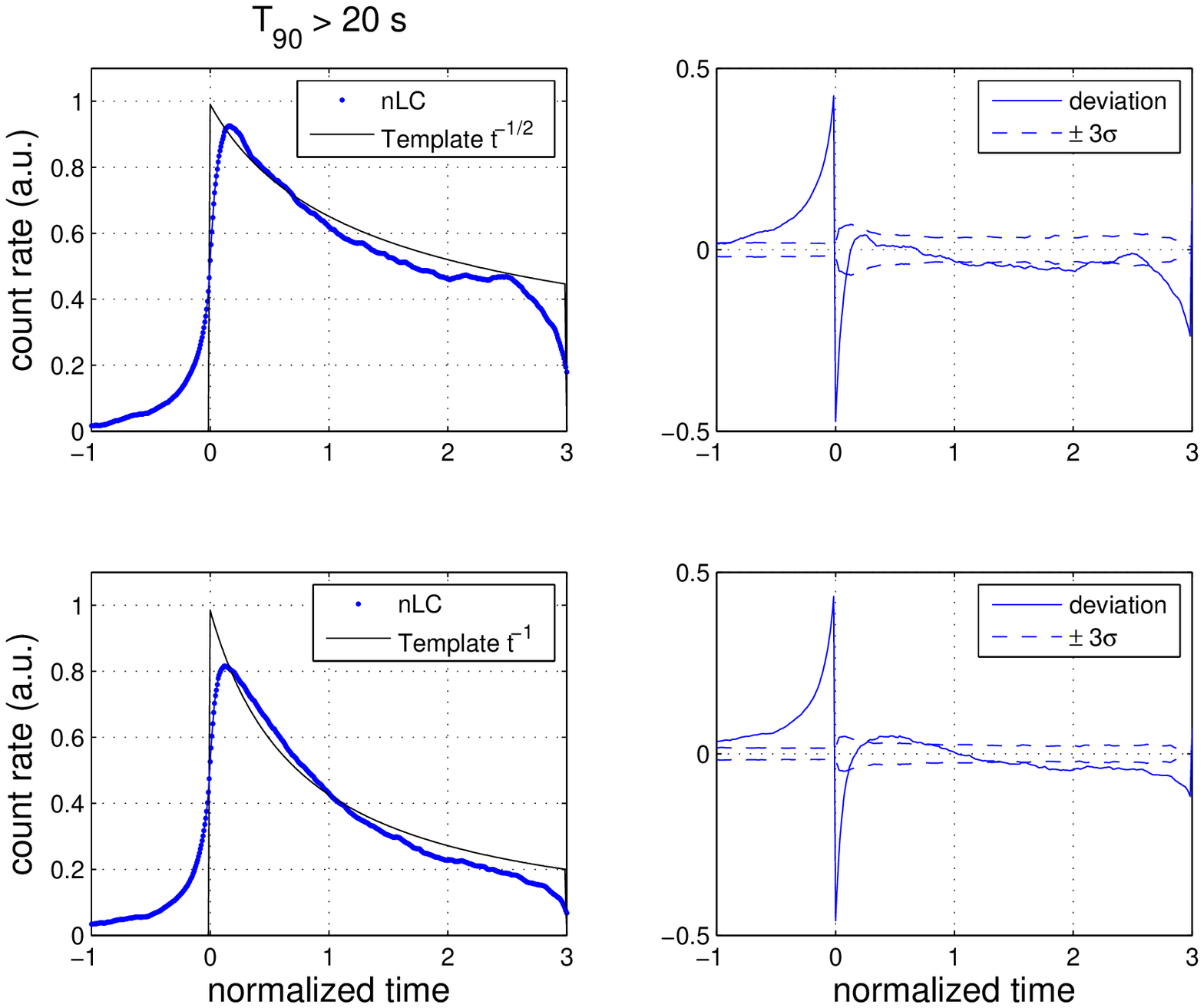}}
	\caption{Phase II matched filtering analysis of BATSE against accretion profiles (\ref{EQN_LII}). Shown are normalized light curves (thick lines) extracted from the BATSE catalog and model llight curves (thin lines) for accretion rates $\propto t^{-n}$ at early time high accretion rates ($n=\frac{1}{2}$) or at low accretion rates ($n=1$) (top panels) and $n=2.5$ and $n=2.75$ (lower panels). All results show a mismatch to the prompt switch-on in model templates.}
	\label{FIG3}
\end{figure*}

\begin{figure*}
	\centerline{\includegraphics[scale=0.47]{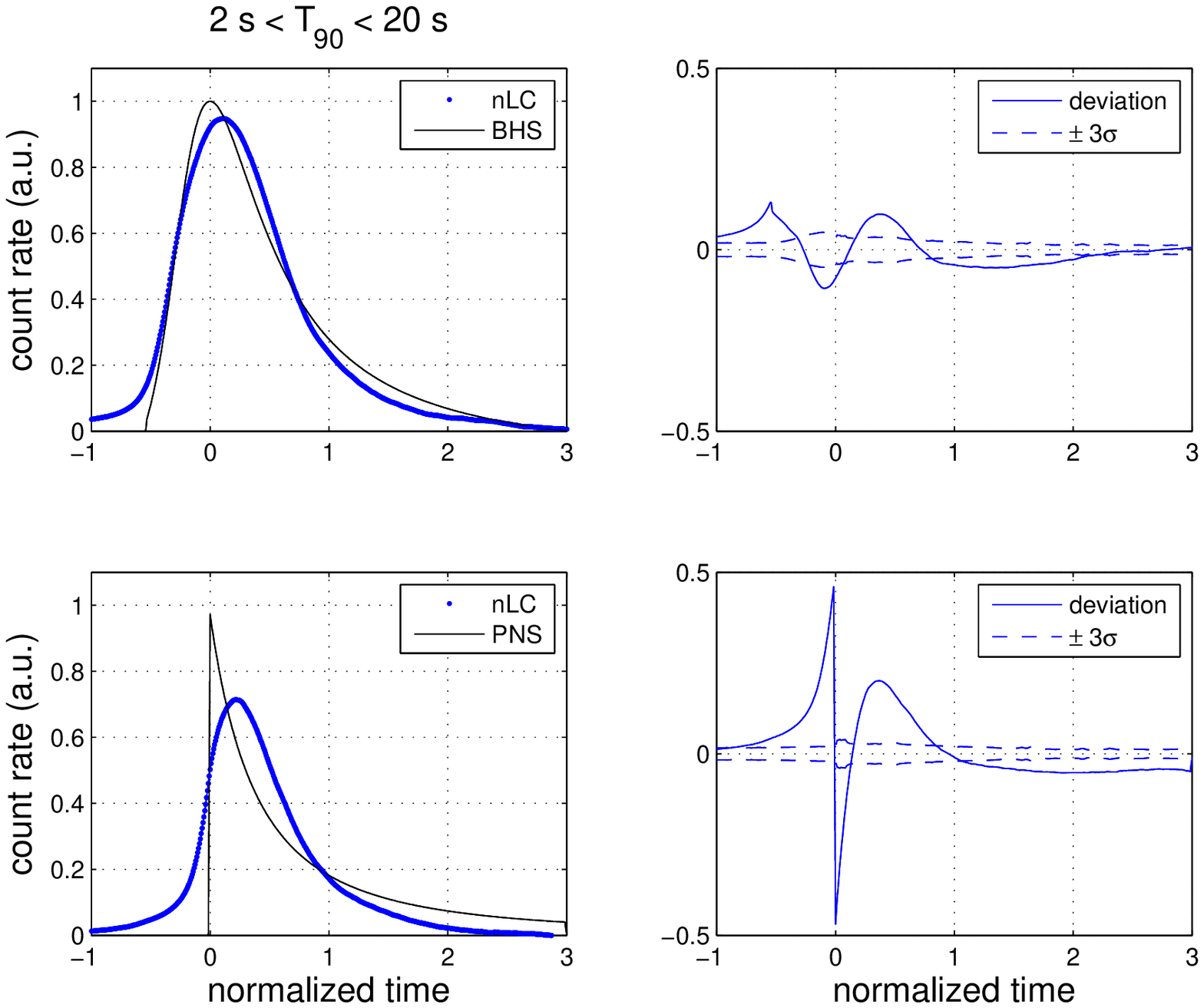}\includegraphics[scale=0.47]{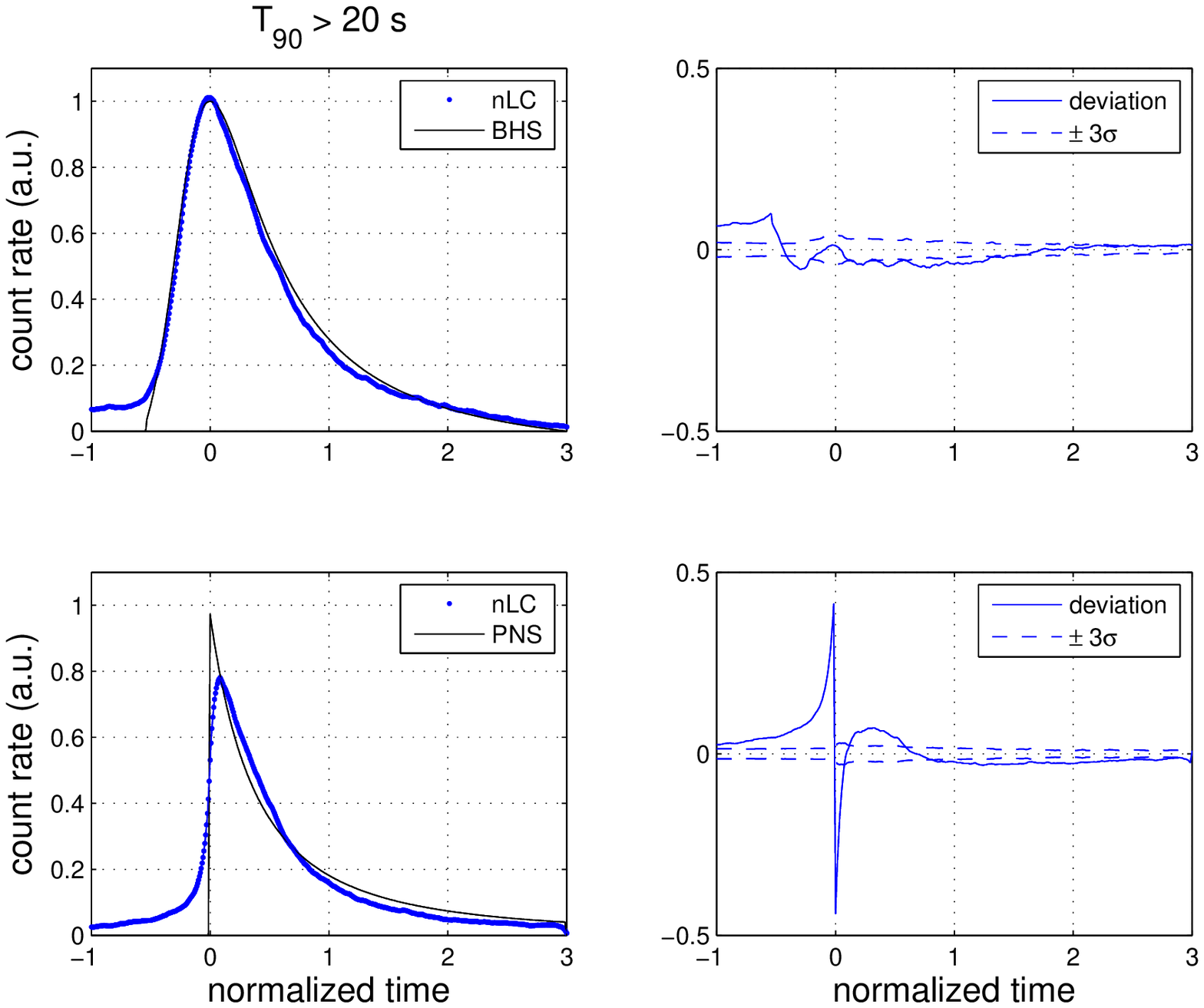}}
	\caption{Phase III matched filtering analysis of BATSE against the model light curve of Fig. \ref{FIG2} (top panels) and that of a proto-neutron star (PNS, lower panels). Consistency is relatively better for the former, especially so for durations greater than 20 s. We attribute this time scale to that of jet breakout of a stellar remnant envelope. 
	(Adapted from \citep{van12}.)
	}
	\label{FIG4}
\end{figure*}

Our light curve of Fig. 2 has been vigorously confronted with data from
BATSE, {\em BeppoSAX}, {\em Swift} and HETE-II. We mention the following.
\begin{enumerate}
	\item {\em Long durations.} $T_{spin}$ in (\ref{EQN_Tspin}) defines a secular time scale of tens of seconds for poloidal magnetic field energies at the limit of stability, defined by the kinetic energy in a torus at the ISCO \citep{van99,van03}. For LGRBs, $T_{90}\simeq T_{sp}$ of initially rapidly rotating black holes with $\sigma\simeq1\%$. It also gives an improved correlation $E_\gamma \propto T_{90}^\beta E_p$ $\beta\simeq 0.5$) between the true energy in gamma-rays $E_\gamma$ (corrected for beaming) and the peak energy $E_p$ in gamma-rays \citep{van08b,sha15}. 
	\item {\em Universality.} The dichotomy of long and short GRBs can be identified with hyper- and suspended accretion onto initially slowly and, respectively, rapidly spinning black holes \citep{van01a}. It implies various common emission features. First, SGRBs are expected to produce X-ray afterglows similar to LGRBs, albeit less luminous in host environments typical for mergers. This prediction \citep{van01a} is confirmed by the {\em Swift} and HETE-II events GRB050509B \citep{geh05} and GRB050709 \citep{fox05,hjo05,vil05}. Second, mergers involving black holes with rapid spin may feature soft Extended Emission (EE) in suspended accretion, that bears out in GRBEEs such as GRB 060614 \citep{del06} satisfying the same Amati relation of long GRBs (recently reviewed in \citep{van14b}). Third, remnants of GRBs are all slowly spinning black holes with $a/M\simeq 0.36$ representing the late-time fixed point $\Omega_H \simeq \Omega_{ISCO}$. With no memory of the initial spin of the black hole it predicts common features from late time accretion or fall back matter. Notably, the {\em Swift} discovery of X-ray tails (XRTs) may result, in mergers, from messy break-up of a neutron star \citep{lee98,lee99,ros07}, e.g., in GRB060614 involving a rapidly rotating black hole \citep{van08a}. 
	\item {\em Spin down in BATSE.} BATSE light curves of LGRBs have been analyzed by matched filtering. Figs. \ref{FIG3}-\ref{FIG4} shows results matched filtering analysis for Phase II and III, given by (\ref{EQN_LII}) and, respectively, Fig. \ref{FIG2}. Included in Fig. \ref{FIG4} is further a light curve from spin down of a magnetar. The {\em normalized light curves} (nLC) extracted from BATSE represent averages of 960 individually normalized BATSE light curves with $T_{90}> 20$ s and 531 light curves with $T_{90}<20$ s. The nLC attains a peak at about 16\% of its duration. The different matches in Figs. \ref{FIG3}-\ref{FIG4} show that long GRBs have a gradual switch-on as in spin down away from the unstable fixed point $\Omega_H=\Omega_T=1/2M$ in (\ref{EQN_CC}), rather than prompt switch-on implied by (\ref{EQN_LII}) or the light curve of a newly formed magnetar. Also, very long duration events ($T_{90}>20$ s) show a remarkably good match to the model template of Fig. \ref{FIG2}) for all time. On average, LGRBs appear more likely to derive from a spin down Phase III rather than Phase II or a magnetar. 
	\item {\em No proto-pulsars.} There is a non-detection of proto-pulsars in broadband Kolmogorov spectra of the 2kHz {\em BeppoSAX} spectra of long GRBs, extracted by a novel butterfly fliter by matched filtering against a large number of chirp templates \citep{van14a}. 
	\item {\em Bright is variable.} LGRBs show a positive correlation between luminosity and variability \citep{rei01}. In our model, black hole feedback on accretion flow provides a novel channel for instabilities that gives rise to a luminosity proportional to the inverse of the duty cicle \citep{van15a}.
\end{enumerate}

We next confront our model with the light curve of ASASSN-15lh.


\section{Confrontation with SN2015L}

\begin{figure*}[h]
	\centerline{\includegraphics[scale=0.7]{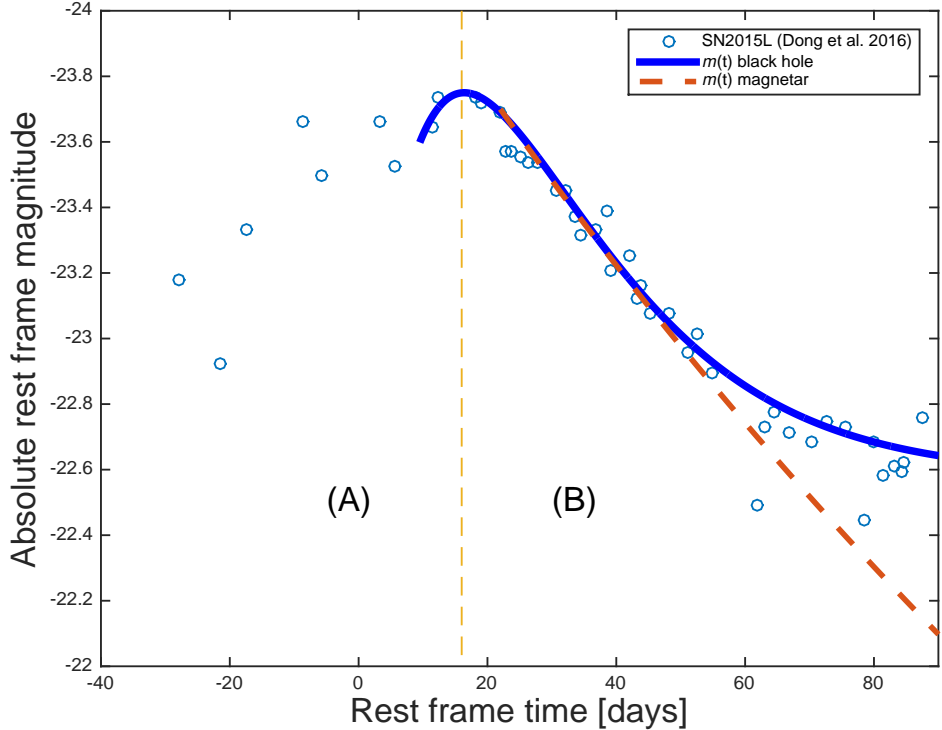}}
	\caption{Shown is a confrontation of the model light curve of black hole spin down in Phase III and that of magnetars to the bolometric light curve of SN2015L (Table S2 in \citep{don16}) by scaling of peak magnitude and duration (B). The theoretical change in magnitude $m_{pl}-m_p=1.15$ post-peak agrees with the observed change $\Delta m\simeq 1.2$. Prior to peak luminosity (A), a near-extremal black hole forms during a Bardeen accretion Phase II.}
	\label{FIG5}
\end{figure*}

Fig. \ref{FIG5} shows a match to the bolometric luminosity of SN2015L of our light curve $L_{BPJ})(t)$ with dissipation largely into heat in the optically thick remnant stellar envelope. On this basis, we identify SN2015L with the relaxation of a Kerr spacetime to that of a slowly spinning black hole, featuring a late time plateau $\Omega_H\simeq \Omega_{T}<<1/2M$. According to Fig. 2, the magnitude of $L_{BPJ}$ rises by about one magnitude post-peak peak. A prior onset to peak is associated with the formation of a rapidly rotating black hole in prior epoch of hyper-accretion \citep{van15b}.

We emphasize that {only {\em one} scale factor (\ref{EQN_Ts2}) is applied to $L_{BPJ}(t)$ to match SN2015L}, here scaled down by (\ref{EQN_alpha}) from the same applied to long GRBs.
The predicted transition post-peak to a plateau is specific to spindown of black holes with no counterpart to magnetars, since the latter continue to spin down freely to essentially zero angular velocity at late times \citep[e.g.][]{dai16}. Our value (\ref{EQN_Dm}) is in remarkable agreement with the observed change $\Delta m =1.2$. 

A 120-day UV rebrightening to SN2015L 90 days after its onset \citep{god16} is a further striking feature in SN2015L. We interpret it as late-time activity of the inner engine, again by virtue of aforementioned short light crossing and cooling time scale of the expanding envelope. In our model, the remnant inner engine of the 90-day ``prompt" SN2015L light curve is a slowly spinning black hole, in common with the remnant of SGRBs and LGRBs. Any late time accretion on these engines produces a latent emission at a luminosity essentially set by the accretion rate. In the case of GRBs, we hereby explain the XRTs, common to both short and long GRBs discovered by {\em Swift}. In the case of SN2015L, we attribute the UV rebrightening analogously, by late time accretion onto the slowly rotating remnant. A further illustration hereof is the decade-long X-ray emission in the supernova remnant of SN1979C by accretion onto a remant black hole \citep{pat11}. According to the Bardeen accretion Phase II, this can lead to spin up, whereby the plateau - XRT in case of GRBs and UV rebrightening in case of SN2015 - may fluctuate slightly in luminosity.  

\section{Conclusions and outlook}

SN2015L presents a significant addition to the class of extreme transient events, typically
associated with core-collapse supernovae and GRBs. By its large amount of radiation $E_{rad}$ satisfying (\ref{EQN_rk}) with a light curve featuring a late time plateau satisfying (\ref{EQN_Dpl}), SN2015L may present a major new class of SLSNe. If so, we expect to see even brighter events in future surveys. 

Core-collapse of massive stars are believed to be factories of neutron stars and black holes, that may be powering aspherical explosions by outflows derived from their energy reservoir in angular momentum. These alternatives give rise to different model light curves in electromagnetic radiation from dissipation of magnetized outflows, further in light of different phases in black hole evolution
in three consecutive steps shown in Fig. 2.

As universal inner engines, black hole outflows can be scaled in time for consideration to SN2015L based on (\ref{EQN_rg}) and LGRBs according to (\ref{EQN_alpha}). In our model, the light curve of SN2015L and prompt GRB-emission are associated with black hole spin down. Durations hereof are defined by the lifetime of spin (\ref{EQN_Tspin}). Scaling of $T_s$ is represented by different ratios $\sigma$ of torus to black hole mass. Our model hereby contains essentially one parameter, assuming stellar mass black hole masses to vary by at most a factor of a few.

In light of scaling by $\sigma$, we confront our theoretical model light curves with both LGRBs from BATSE and SN2015L. Results on normalized light curves shows satisfactory agreement with black holes loosing angular momentum to matter at the ISCO, more so that a prior accretion powered phase or spin down of magnetars.

The model light curve of Fig. \ref{FIG2} shows a satisfactory match to the bolometric luminosity light curve of SN2015L, here attributed to effective dissipation in the remnant stellar envelope. The late-time plateau is identified with gradual spin down of an initially near-extremal Kerr black hole to a slowly rotating black hole, whose angular velocity has settled down to that of matter at the ISCO. This scenario is exactly the same as identified for LGRBs in BASTSE. The late-time state of $\Omega_H\simeq \Omega_{ISCO}$ defines a plateau in the light curve unique to black hole-torus system, which is absent in magnetars. (Their spin decays all the way to zero.) Quantitative agreement is found in the change in magnitude $\Delta m$ post-peak in our model light curve and the light curve of SN2015L.

It appears that SN2015L is genuinely powered by the spin energy of a rotating black hole interacting by frame dragging induced Alfv\'en waves with surrounding matter at the ISCO, delivering a total energy output typical for normal long GRBs (\ref{EQN_rg}) defined by $E_{rot}$ of stellar mass near-extremal black holes. The long duration of months is here identified with the lifetime of black hole spin, subject to spin down against relatively low density accretion flow. The commensurably lower luminosity is readily radiated off by the envelope in optical emission, whereby the BPJ from the black hole fails to reach successful stellar break-out. 

{A principle outcome of the present model is an accompanying major output $E_{GW}>>E_{rad}$ in gravitational waves and a slowly rotating black hole remnant, where $E_{GW}$ is emitted over the course of spin down (tens of seconds for LGRBs, months for SN2015L type events) and the final remnant satisfies {a relatively high mass low spin black hole} (Fig. \ref{FIG1b})
	\begin{eqnarray}
	M\simeq M_0,~~a/M\simeq 0.36,
	\label{EQN_C}
	\end{eqnarray}
	{defined by the outcome of black hole growth in spin-up by Bardeen accretion and subsequent spin-down to the stable fixed point $\Omega_T=\Omega_H$ (in the approximation $\Omega_T\simeq \Omega_{ISCO}$). This outcome should be contrasted with moderate mass black holes at slow spin produced by direct accretion alone, immediately following black hole birth.} {The outcome (\ref{EQN_C}) explains the estimated mass and spin parameters of the recent black hole binary merger GRB150914 \citep{lig16}}
	\begin{eqnarray}
	M_1=35.7_{-3.8}^{5.4}M_\odot,~~a_1/M_1=0.31^{+0.48}_{-0.28},\\
	M_2=29.1_{-4.4}^{3.8}M_\odot,~~a_2/M_2=0.46^{+0.48}_{-0.42}
	\end{eqnarray}
	{\em We speculate that the progenitor binary of GWB1509 is a merger of SN2015L type remnants from stars of mass $M_0\le 50M_\odot.$} 
		
	{The final spin down evolution leading to (\ref{EQN_C}) represents the 
	liberation of an appreciable fraction of black hole spin energy $E_{rot}$ into 
	and output $E_{GW}$ in gravitational waves. According to Fig. 3, the efficiency $\eta$ of converting $E_{rot}$ to radiation can reach up to 60\%. Since $E_{GW}$ is expected to be the dominant output channel (over, e.g., magnetic winds \citep{van03}; in the application to long GRBs, the latter further includes MeV neutrino emission in a time scale of tens of seconds), an energetic output
	\begin{eqnarray}
	E_{GW}\simeq 2\times 10^{54}\left(\frac{\eta}{50\%}\right) \left(\frac{M}{10M_\odot}\right) \,\mbox{erg}
	\end{eqnarray}
	from spin down of rapidly rotating stellar mass Kerr black holes in core-collapse such as described in Figs. 2-3 \citep{van16}. Emitted during spin down, this creates a descending chirp in the time-frequency domain, by expansion of the ISCO during black hole spin down. At late time, it assymptotes to a late time quadrupole frequency indicated in Fig. \ref{FIG1c}, satisfying
    \citep{van11a}
	\begin{eqnarray}
	f_{GW,3} \simeq 600 - 700 \,\mbox{Hz}\,\left(10M_\odot/M\right).
	\label{EQN_fGW}
	\end{eqnarray}
    Continuing emission may ensue by late time accretion in a subsequent plateau at these frequencies, defined by the stable fixed point $\Omega_H = \Omega_{ISCO}$. If detected, (\ref{EQN_fGW}) provides a rigorous measurement of the mass of the putative black hole, that may be searched for by chirp-based spectrograms \citep{van16}.}
	
{Our scaling to superluminous SNe of (\ref{EQN_fGW}) originally developed for long GRBs points to simular frequencies at much longer durations up to the time scale of months revealed by the light curve of SN2015L shown in Fig. \ref{FIG5}.
Based on \cite{qui13}, we anticipate an approximate event rate of a few such extreme
SN2015L type events Gpc$^{-3}$ yr$^{-1}$, i.e., up to a dozen of SLSN-I per year within a few hundred Mpc. It therefore seems worthwhile to pursue a multimessenger view on these remarkable events.}

\section{Supporting information}

Additional supporting information may be found in the online version of this article:\\ \\
{\bf BHGROWTH}: MatLab program of Figs. \ref{FIG1a}-\ref{FIG1c}.\\ 
{\bf BATSENLC:} Fortran program of BATSE analysis in Fig. \ref{FIG4}.\\

{\bf Acknowledgments.} The authors thank A. Levinson and D. Eardley for stimulating discussions and the referee for her/his constructive comments on the manuscript. MVP thanks the Kavli Instiute for Theoretical Physics, UCSB, were some of the work has been performed. 
BATSE data are from the NASA GRO archive at Goddard.
This research NSF-KITP-16-015 was supported in part by the National Research Foundation of Korea  (2015R1D1A1A01059793, 2016R1A5A1013277) and the National Science Foundation under Grant No. NSF PHY11-25915.


\bsp	
\label{lastpage}
\end{document}